\documentclass[prb,twocolumn,showpacs,preprintnumbers,amsmath,amssymb]{revtex4}

\usepackage{graphicx}
\usepackage{dcolumn}
\usepackage{bm}

\begin{document}

\preprint{APS/123-QED}

\title{Fiske steps studied with flux-flow resistance oscillation in a narrow stack of Bi$_{{\rm 2}}$Sr$_{{\rm
2}}$CaCu$_{{\rm 2}}$O$_{{\rm 8}{\rm
+} {\rm d}}$ junctions}

\author{S. M. Kim, H. B. Wang, T. Hatano, S. Urayama, S. Kawakami, M. Nagao, Y. Takano, T. Yamashita}
\affiliation{%
National Institute for Materials Science (NIMS), 1-2-1 Sengen, Tsukuba 305-0047 Japan\\
}%

\author{K. Lee}
\affiliation{ Department of Physics and Interdisciplinary Program
of Integrated
Biotechnology, Sogang University, Seoul 121-742, Korea  \\
}%

PaperId: cond-mat/0412690

\begin{abstract}
We have experimentally investigated the fluxon dynamics in a narrow
Bi$_{\rm 2}$Sr$_{\rm 2}$CaCu$_{\rm 2}$O$_{\rm 8+d}$ stack with
junction length $L\sim $1.8 $\mu $m. As an evidence of
high-frequency excitation by a collective cavity mode, under an
(in-plane) external magnetic field, the current-voltage
characteristics show prominent Fiske steps with the corresponding
resonance frequencies of 75-305 GHz. Further study of flux-flow
resistance oscillation with various $c$-axis currents clarifies the
correlation with Fiske steps by distinguishing two different regions
i.e., static flux-flow region at low bias current level and dynamic
Fiske step region at high bias current level.
\end{abstract}

\maketitle

\begin{figure}
\includegraphics[width=6cm]{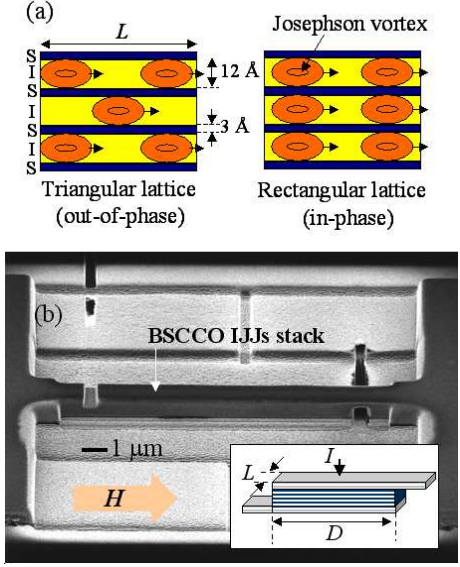}
\caption{\label{fig.1}(Color online) (a) The configurations of
Josephson vortex lattice in a stacked BSCCO IJJs. The triangular
lattice corresponds to the out-of-phase case, while the
rectangular lattice corresponds to the in-phase situation. (b)
Scanning ion-beam microscope image of BSCCO stack fabricated by
the focused ion beam etching. Schematic diagram of the stack is
shown in the inset. Magnetic field $H$ was applied along the
longer side ({\textit{D}}) of junction to enhance edge effect.}
\end{figure}

In stacked Bi$_{{\rm 2}}$Sr$_{{\rm 2}}$CaCu$_{{\rm 2}}$O$_{{\rm
8}{\rm +} {\rm d}}$ (BSCCO) intrinsic Josephson junctions (IJJs),
the mutual interaction between junctions is expected since the
superconducting layers, with the layer thickness $d$ = 0.3 nm, are
much thinner than the London penetration depth \textit{$\lambda
$}$_{L}$= 170 nm\cite{Sakai:1998}. Under the applied magnetic field
parallel to the layers, Josephson vortices in a stack form lattice
configurations that depend on a constant phase shift between
neighboring layers, ranging from 0 for the rectangular lattice to
$\pi$ for the triangular lattice (see Fig.~\ref{fig.1}(a))
\cite{Bulaevskii:1991,Koshelev:2001,Kim:2002}.

Under the appropriate conditions, the fluxon lattice will excite the
two-dimensional cavity modes in $N$ stacked junctions, leading to
the emission of electromagnetic wave with characteristic frequency
$f_{nm} \approx mc_{n} / 2L$ ($c_{n} $ is the phase velocity of
electromagnetic wave, $L$ is the junction length, $m$ (=1, 2,
3,\ldots) denotes $L$-direction mode, and $n$ (=1, 2, 3,\ldots $N$)
denotes the stacking direction mode)\cite{Kleiner:1994}. The $c_{n}
$ is given by
\begin{eqnarray}
 c_{n} = \omega _{pl} \lambda _{J} [1 - 2S\cos (\pi n / (N + 1))]^{ - 1 /
2} , \label{eq:1}
\end{eqnarray}
\noindent with the Josephson penetration depth $\lambda _{J} $, the
coupling parameter $S$, the junction number $N$ and the Josephson
plasma frequency $\omega _{pl}$\cite{Kleiner:1994, Sakai:1998,
Bulaevskii:1991}. Among the $N$ different modes along the stacking
direction, the mode with the lowest velocity $c_{N} $ is usually
stimulated by a triangular vortex lattice; while for the mode with
the highest velocity $c_{1} $, a rectangular lattice is
favorable\cite{Ustinov:1996, Pedersen:1995}.

Being driven by a $c$-axis bias current, the moving fluxon lattice
generates flux-flow resistance (FFR) in the junctions. Recent
experimental\cite{Ooi:2002, Kakeya:2004, Hatano:2005} and
numerical\cite{Machida:2003, Koshelev:2002} studies show that the
motion of traveling fluxon lattice in the layers is reflected into
the periodic oscillation of FFR under the low bias current by
considering the dynamical matching between the vortex lattice and
the sample edges. The $H_{0}/2$-period oscillation of FFR is
interpreted as a result of the formation of triangular lattice in a
long-junction stack, where $H_{0}$ is the field for adding
\textit{one} flux quantum per \textit{one} junction ($H_{0}\equiv
\Phi _{0} / Ls$, $\Phi _{0}^{} $ being the flux quantum and $s$ the
layer periodicity along the $c-$axis\textit{} i.e. 1.5
nm)\cite{Ooi:2002, Machida:2003}. Recent studies indicate that when
the junction size is reduced down to a few $\mu$m and approaches the
short-junction stack limit ($L<\lambda _{J}$), the oscillation
period becomes predominantly $H_{0}$ rather than $H_{0}/2$ due to
the deformation of the Josephson vortex lattice by strong
interaction with junction edges\cite{Kakeya:2004, Hatano:2005}. The
result may suggest a possible existence of collective vortex motion
such as a rectangular vortex lattice in a narrow stack. However the
recent FFR research is limited at very low bias current levels such
as less than 1\% of critical current at zero field $I_{c0}$, and it
is still obscure whether the vortex dynamics studied by FFR is
related with high-frequency excitation or not.

In this paper, stimulated by the above-mentioned researches, we
confine ourselves to study of the singularities in the $I-V$ curves
and the correlation with the current-dependent FFR oscillation in
the narrow BSCCO stacks with $L\sim $1.8 $\mu $m and large junction
number $N > > 1$, exploring the high-frequency excitation by a
possible collective fluxon motion.

In our experiments, a BSCCO whisker\cite{Nagao:2001} with a clean
and flat surface was fabricated to be an \textit{in-line} type of
IJJs stack with the length of 1.8 $\mu$m by a focused ion beam. A
schematic diagram of the stack is shown in the inset of
Fig.~\ref{fig.1}(b). Here, $L$ and $D$ denote the junction length
perpendicular to the magnetic field and the depth parallel to the
field. The fabricated IJJs stack had a thickness of about 180 nm,
i.e., containing about 120 junctions. The values of critical
temperature ($T_{c} $)\textit{} was 81 K, and the values of $I_{c0}
$ were about 350 $\mu $A at 10 K and about 195 $\mu $A at 50 K
respectively. The electric transport properties were measured with a
four-terminal configuration using a Physical Property Measurement
System (PPMS, Quantum Design), which can supply magnetic field up to
9 T. In order to enhance the edge effect of the sample on moving
Josephson vortices, we applied the magnetic field parallel to the
\textit{ab}-plane along the longer side ($D$) of the BSCCO stack
(see Fig.~\ref{fig.1}(b)). The sample was mounted on a rotatable
holder with a resolution better than 0.005$^{{\rm o}}$. The
\textit{in}-plane alignment was precisely adjusted by the angular
dependence of FFR under external magnetic field.

The Josephson penetration depth $\lambda _{J} $ is given by $\lambda
_{J} = [\Phi _{0} / 2\pi \mu _{0} j_{c} (t_{eff} + 2\lambda ^{2} /
d_{eff} )]^{1 / 2}$ with the effective values $d_{eff} = \lambda
\sinh (d / \lambda )$, $t_{eff} = t + 2\lambda \tanh (d / 2\lambda
)$, the (in-plane) magnetic penetration depth $\lambda (T) = \lambda
_{ab} (0) / \sqrt {1 - (T / T_{c} )^{4}} $, by assuming $\lambda
_{ab} (0)$=170 nm, the thickness of superconducting layers and
insulating layers $d$=0.3 nm and $t$=1.2 nm\cite{Hechtfischer:1997}.
The $\lambda _{J} $ of our sample was calculated as 0.27 $\mu $m at
10 K (0.34 $\mu $m at 50 K). The sample width of 1.8 $\mu $m is
about 6.6 times (about 5.3 times at 50 K) as large as the calculated
$\lambda _{J} $ at 10 K, thus it can be still regarded as a
long-junction stack; however, as we applied magnetic fields
perpendicular to the narrower side, which is different from the
conventional case, we use the term `narrow' to specify our sample.

\begin{figure}
\includegraphics [width=7cm]{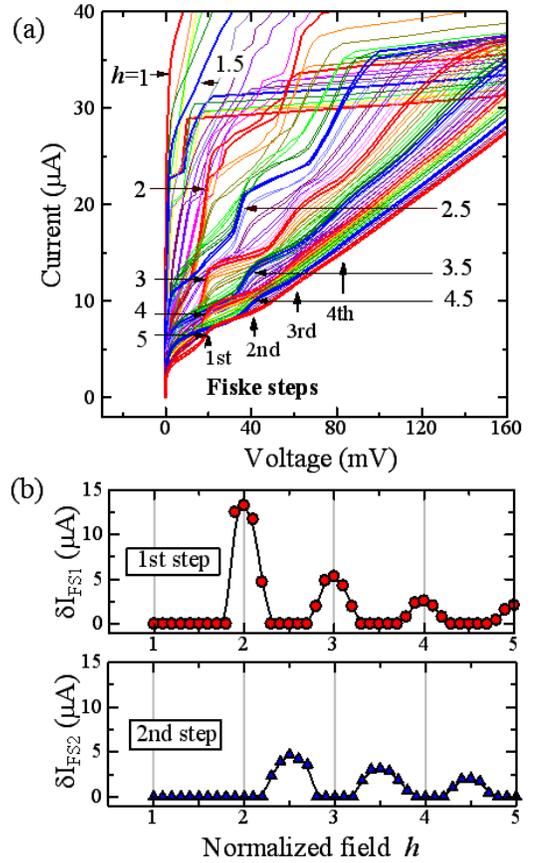}
\caption{\label{fig.2}(Color online) (a) $I-V$ characteristics at
various fields from 0.765 T ($h = 1$) to 3.825 T ($h = 5$) with an
interval of 0.0765 T (0.1 $h$) at 50 K. Bold lines show $I-V$ curves
at fields $h$= integer and half integer where Fiske steps are
clearly observed. (b) The magnetic field dependence of the current
amplitudes of first step and second step at 50 K.}
\end{figure}

Figure~\ref{fig.2}(a) displays the $I-V$ characteristics under
various magnetic fields parallel to the layers at 50 K. In order to
indicate a periodic modulation of the $I-V$ curves with field, we
normalized the magnetic fields in graphs of Fig. 2 by the period
$H_{P} $=0.765 T, which corresponds to $H_{0} $ calculated for this
sample. Then we can regard the normalized field $h \equiv H / H_{P}
( = H / H_{0} )$ as Josephson fluxon number per unit junction. The
magnetic field was applied in the range from 0.765 T ($h = 1$) to
3.825 T ($h = 5$) with an interval of 0.0765 T (0.1$h$).

With increasing magnetic field, we found clear current steps that
were regularly developed in the $I-V$ curves, as shown in
Fig.~\ref{fig.2}(a). These steps were identified as Fiske steps that
had been observed as a strong enhancement of superconducting current
when the Josephson frequency ($\omega = 2eV / \hbar $) matches the
resonant frequency of electromagnetic cavity modes excited in
junctions\cite{Kulik:1964, Cirillo:1998, Krasnov:1999,
Latyshev:2002}. The asymptotic voltage positions of step series are
given by
\begin{eqnarray}
 V = mN(\Phi _{0} c_{n} / 2L). \label{eq:2}
\end{eqnarray}

Using the voltage position of the first Fiske step (18.7 mV) defined
by local maximum $d$I/$d$V, the characteristic velocity was
estimated to be $c_{n} = 2.71\times 10^{5}$ m/sec using
Eq.~(\ref{eq:2}) with $m$=1, $N$=120, and $L$=1.8 $\mu$m. Since the
series of Fiske steps are observable up to the 4th order in
Fig.~\ref{fig.2}(a), the corresponding resonance frequency lies in
the range of 75-305 GHz.

According to Eq.~(\ref{eq:1}), using a set of the experimental
parameters (the current density $j_{c}$=1.015 kA/cm$^{{\rm 2}}$ at
50 K, the thickness of insulating layer $t$=1.2 nm, the coupling
constant $S \approx 0.5$, and $\lambda _{J} $=0.76 $\mu $m), the
velocities of the lowest mode and the highest mode are calculated as
$c_{120} = 2.62\times 10^{5}$ m/sec and $c_{1} = 1.86\times 10^{7}$
m/sec, respectively. Our result, estimated from the Fiske step, is
comparable with the velocity of the lowest collective cavity
resonance mode, in agreement with Refs. \onlinecite{Krasnov:1999}
and \onlinecite{Latyshev:2002}.

The observed odd and even steps obviously have different dependences
on the magnetic field. Fig.~\ref{fig.2}(b) shows the field
dependence of the step height ($\delta I_{FS}$) for the first and
second steps. With increasing field, each height of the first and
the second order steps oscillated with the same period $H_{0}$.
However, the maxima of the first order step appeared at $h$=integer,
and the minima at $h$=half-integer; while the second order step
exhibited opposite behavior. Note that this is very similar to the
behavior of a single junction\cite{Kulik:1964, Cirillo:1998} except
that large current-step voltage due to the contribution from all
junctions. This is very important because such a single junction
behavior is possible when all junctions are evenly excited by a
collective cavity resonance in stacked junctions.

For further understanding of Fiske steps, we also studied the
correlation between Fiske steps and the FFR oscillation as function
of magnetic field, which has been regarded as a powerful tool for
this purpose\cite{Ooi:2002, Kakeya:2004, Hatano:2005, Machida:2003,
Koshelev:2002}.

\begin{figure}
\includegraphics[width=7cm]{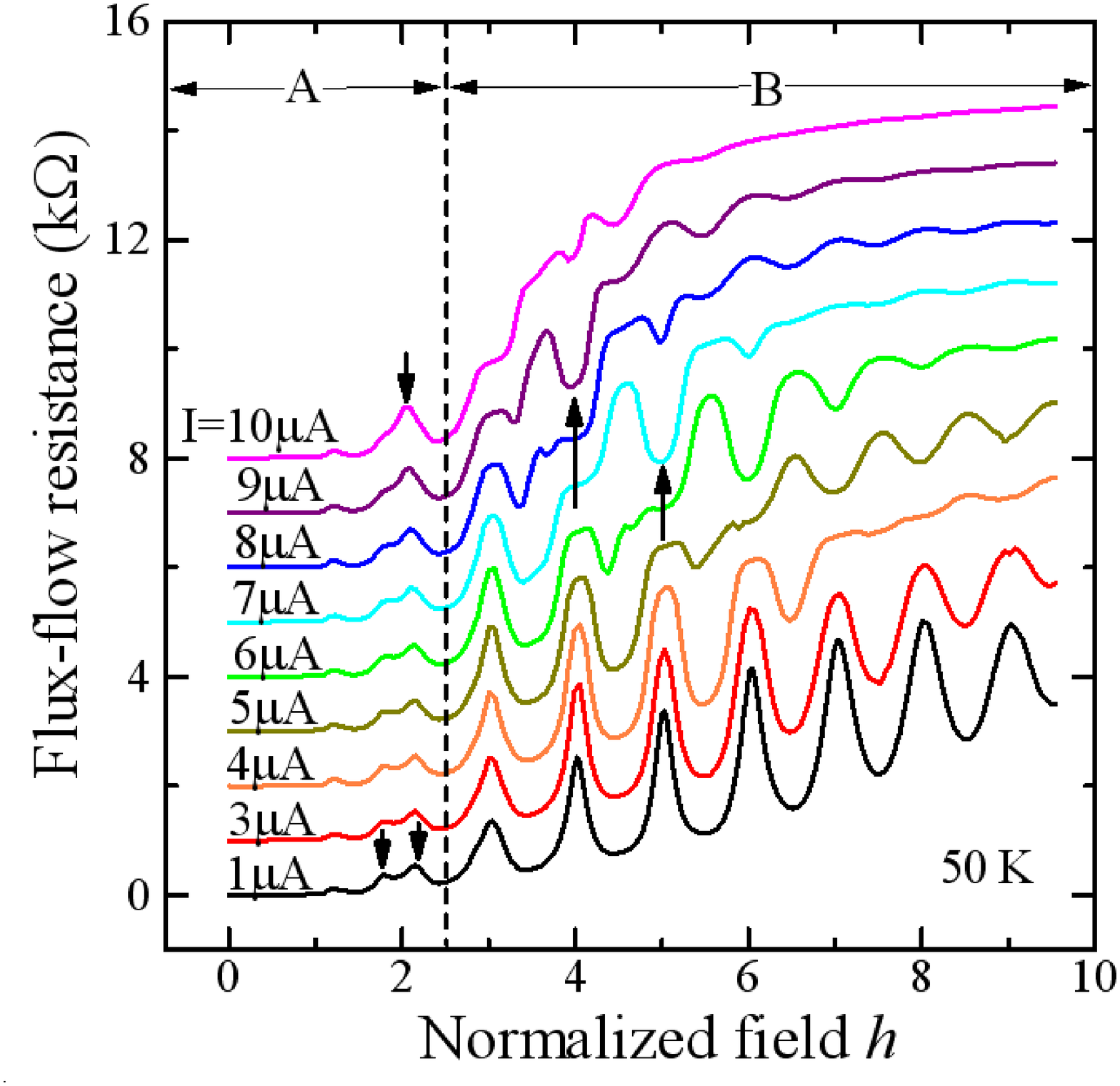}
\caption{\label{fig.3}(Color online) FFR anomalies due to $c$-axis
bias current of 1-10 $\mu $A at 50 K, where each curve is shifted by
1 k$\Omega $ for clarity. The magnetic fields are normalized by
$H_{p} = 0.765$ T. With increasing current, there are two kinds of
features; Part A: two small peaks transform to one peak with period
$H_{0} $ at $h < 2.5$. Part B: the inversion from peaks to local
minimums at $h > 2.5$. The dash dot line in Part B shows the
boundary of two different regimes with $c$-axis bias current.}
\end{figure}

At extremely low $c$-axis current $I = 1~\mu $A (0.52 \% of $I_{c0}
$), as shown in Fig.~\ref{fig.3}, there are two oscillation regimes
of FFR: one is below $h = 2.5$ and the other is over $h = 2.5$ with
the oscillation period of $H_{0} $ (see Part A and B in
Fig.~\ref{fig.3}). The details for the change in oscillation period
by magnetic field will be elucidated with the model based on the
edge current elsewhere\cite{Hatano:2005}.

Having noticed in Fig.~\ref{fig.2}(a), there is not any step at such
a low current level, we intended to measure the FFR at much high
current levels. Surprisingly, we have observed a new anomaly of FFR
oscillation at higher bias currents as shown in Fig.~\ref{fig.3}. In
particular, for high field $h > 2.5$ (Part B in Fig.~\ref{fig.3}) it
is found that the peaks and the local minima of FFR oscillation were
inverted by varying bias current as marked by the arrows, with the
oscillation period $H_{0} $ unchanged. Note that the inversion from
the peaks to the local minima of FFR oscillation implies a change of
the matching between the edge and Josephson vortices as well as an
enhanced conductance. As is well known, such conductance enhancement
in a Josephson junction, usually resulting from resonance in the
junction, should manifest itself as a current steps in the $I-V$
characteristics.

To confirm this, we compared the current-dependent anomaly of FFR
oscillation in Fig.~\ref{fig.3} with the $I-V$ curves in
Fig.~\ref{fig.2}(a). Looking at the FFR curves marked by the arrows
at the normalized field $h = 4$ and 5 in Fig.~\ref{fig.3}, the clear
inversion from peaks to local minima took place when the bias
current was varied from 5 $\mu $A to 9 $\mu $A. Considering the
corresponding curves in Fig.~\ref{fig.2}(a) with the same field and
bias current range, we see two Fiske steps appearing at the same
voltage in the $I-V$ curves, as marked by the arrows at the fields
$h = 4$ and 5. As one can see in Fig.~\ref{fig.3}, there are several
more inversion points at different normalized fields, and the
corresponding steps in Fig.~\ref{fig.2}(a).

Different from the high field regime $h > 2.5$ where the FFR
oscillates with the period $H_{0}$, in the low field regime $h <
2.5$ where the FFR oscillation period is smaller than $H_{0}$ and
comparable with $H_{0} / 2$, there is neither inversion of the FFR
oscillation nor steps up to 10 $\mu $A. In the case of the field of
$h = 2$ in Fig.~\ref{fig.2}(a), the first Fiske step appears at the
current around 20 $\mu $A. The experimental results show the
inversion of FFR oscillation near 20 $\mu $A after FFR becomes
$H_{0}$-oscillation, i.e., two peaks merge into one peak (see the
arrows in Part A of Fig.~\ref{fig.3}). Accordingly, it is clear that
the Fiske steps appear only in the regime that FFR oscillates with a
period of $H_{0} $ and that there is the inversion of FFR
oscillation from peaks to local minima.

Noticeably, after the inversion of FFR oscillation, there is
distortion of $H_{0} $-oscillation with $I$=8 $\mu $A around $h$=4
and $h$=4.5 where the first and the second steps appear in $I-V$
curves. This can be understood with the following explanation. When
the first and the second Fiske steps coexist at same current level,
FFR will reflect the behavior of these Fiske steps under magnetic
field. Then the product of two (odd and even step's) field
dependence factors (refer to Fig.~\ref{fig.2} (b)) should
consequently result in $H_{0}/2$-period oscillation of FFR. This is
in fact the case at which Ustinov and Pedersen could observe
$H_{0}/2$-oscillation in their simulation for a single
long-junction\cite{Ustinov:2005}. However in our experiments, some
steps don't coexist at the same current due to large voltage
intervals between Fiske steps and the reduced step amplitude at 50
K. Therefore the FFR oscillation was mainly affected from one step
(odd or even), with the dominant $H_{0} $-period at the current near
Fiske steps. Nevertheless, the neighboring steps caused observable
distortion of FFR.

Having understood the correlation between the FFR oscillation and
the $I-V$ characteristics, we can distinguish between the two
different regimes divided by dash dot line in Part B of
Fig.~\ref{fig.3}. In the low bias region before the occurrence of
inversion of FFR, i.e., before the appearance of Fiske steps, the
Josephson fluxon dynamics is mainly determined by the edge pinning
effect and the coupling interaction between the layers\cite{
Machida:2003, Koshelev:2002}. Therefore the information of fluxon
lattice can be probed by the FFR oscillation measurement to some
extent. As a matter of fact, such oscillation of FFR in this regime
(especially at the low bias current and high field) is just same as
$I_{c} $ modulation with magnetic field.

On the other hand, in the inversion region of FFR oscillation
accompanying the Fiske steps at relatively high bias current range,
the dynamic resonant fluxon motion is dominant due to the
interaction between the traveling fluxon and the cavity mode
excitation. Thus in the high bias region showing the Fiske steps,
the fluxon lattice can't be determined simply by the measurement of
FFR oscillation. We can only try to figure out the lattice structure
by comparing the characteristic velocity obtained from the Fiske
steps measurement and Eq.~(\ref{eq:1}).

With low-temperature scanning electron microscopy (LTSEM), Quenter
$et$ $al.$\cite{Quenter:1995} also observed similar different
regimes . For the bias points close to origin (in a finite voltage
state without Fiske steps), the LTSEM image showed the static
distribution of the Josephson current in the presence of an external
magnetic field. For the high bias point close to a voltage of Fiske
resonance, a clear standing-wave pattern is observed due to the
superposition of the traveling wave and the reflected waves. These
are consistent with our results.

In summary, the current-voltage characteristics of BSCCO IJJs stacks
with $L\sim $1.8 $\mu $m under an external magnetic field showed
pronounced Fiske steps as an evidence of high-frequency excitation.
The alternative appearance of even and odd Fiske steps resembles the
behavior of a single junction, although there are more than 100
junctions in one stack. The observed Fiske steps and their mode of
the collective cavity resonance suggest that all junctions in such a
narrow stack can be synchronized by the Fiske resonance. Further
measurements on the field-dependent FFR with various $c$-axis
currents clarify the correlation with Fiske steps by distinguishing
two different regions i.e., static flux-flow region at low bias
current level and dynamic Fiske step region at high bias current
level.

The authors would thank N.F. Pedersen, M. Machida, B. Friedman, M.
Tachiki and I. Iguchi for valuable discussion.

\end{document}